# A Reversed Inverse Faraday Effect

Ye Mou, Xingyu Yang, Bruno Gallas, and Mathieu Mivelle*

Sorbonne Université, CNRS, Institut des NanoSciences de Paris, INSP, F-75005 Paris, France

*Corresponding author: mathieu.mivelle@sorbonne-universite.fr

**Abstract**

The inverse Faraday effect is a magneto-optical process allowing the magnetization of matter by an optical excitation carrying a non-zero spin of light. In particular, a right circular polarization generates a magnetization in the direction of light propagation and a left circular polarization in the opposite direction to this propagation. We demonstrate here that by manipulating the spin density of light, i.e., its polarization, in a plasmonic nanostructure, we generate a reversed inverse Faraday effect. A right circular polarization will generate a magnetization in the opposite direction of the light propagation, a left circular polarization in the direction of propagation. Also, we demonstrate that this new physical phenomenon is chiral, generating a strong magnetic field only for one helicity of the light, the opposite helicity producing this effect only for the mirror structure. This new optical concept opens the way to the generation of magnetic fields with unpolarized light, finding application in the ultrafast manipulation of magnetic domains and processes, such as spin precession, spin currents, and waves, magnetic skyrmion or magnetic circular dichroism, with direct applications in data storage and processing technologies.

# 1. Introduction

The inverse Faraday effect (IFE) is a magneto-optical process allowing the magnetization of matter by optical excitation only.[1-3] This magnetization is made possible by the generation of non-linear forces that light exerts on the electrons of the material.[4-7] In particular, in metal, these non-linear forces will generate continuous drift currents at the origin of the IFE.[7,8] If we consider the free electrons of the metal as free-moving charges, these drift currents can be described by the formalism developed by the plasma community.[9,10] In particular, R. Hertel has shown that the expression of the electron drift currents ($J_d$) in a metal is given by the following equation:[7,8]

$$\boldsymbol{J_d} = \frac{1}{2en} Re\left(\left(-\frac{\nabla \cdot (\sigma_\omega \boldsymbol{E})}{i\omega}\right) \cdot (\sigma_\omega \boldsymbol{E})^*\right) \quad (1)$$

With e the charge of the electron (e < 0), n the charge density at rest, $\sigma_\omega$ the dynamic conductivity of the metal, and **E** the optical electric field

From these currents and via the equation of Biot and Savart (Equation (2)), generating a stationary magnetic field (**B**) by optical excitation is possible.

$$\boldsymbol{B} = \frac{\mu_0}{4\pi} \iiint \frac{\boldsymbol{J_d} \times \boldsymbol{r}}{|\boldsymbol{r}|^3} dV \quad (2)$$

Where $\mu_0$ is the vacuum magnetic permeability, dV is the volume element and **r** is the vector from dV to the observation point.

From Equation (1), we can see that to obtain strong drift currents and thus a strong magnetic field by IFE, it is required to create intense electric fields and field divergence. Therefore, due to its ability to manipulate and increase the optical electric field and its gradients, nanoplasmonics is nowadays the only way to obtain strong magnetic fields at the nanoscale, particularly at ultra-short time scales.[11-18] These special characteristics of magnetization have various applications in the fields of magnetic research and technology.[19] Indeed, researchers have been exploring ways to control and study magnetization at very short time and spatial scales,[20-24] using femtosecond lasers, ever since the pioneering work of Beaurepaire et al.[25], intending to control and accelerate current data storage technologies. However, the physical processes involved in this type of interaction are not yet fully understood. Likewise, the transient processes of magnetic interactions, such as spin precession, spin-orbit coupling, and exchange interactions, occur in the femtosecond time scale.[26] Being able to probe and understand these processes using ultrashort pulses of magnetic fields would greatly benefit research activities in magnetism, including Zeeman splitting,[27] magnetic trappings,[28] magnetic skyrmions,[29] magneto-plasmonics,[30] ultrafast magnetic modulations,[31] and magnetic circular dichroism[32] to spin control,[33] spin precession,[34] spin currents,[35] and spin waves.[36]

From Equation (1), we can also see that a necessary condition must be met to generate drift currents at the origin of the IFE. The light incident on the metal must carry some degree of ellipticity so that the product **E.E\*** is not zero. In particular, we note that a right circular or elliptical polarization will generate a magnetic field oriented in the same direction as the propagation of the light. In contrast, a left circular polarization will create it in the opposite direction of propagation (Figure 1).[1-3]

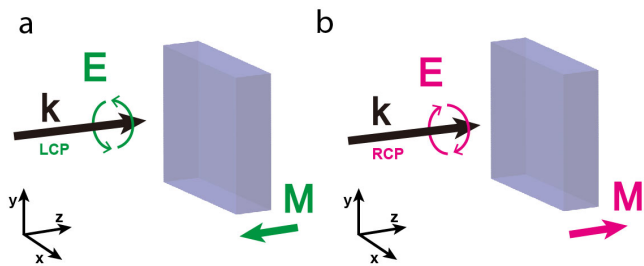

**Figure 1:** The inverse Faraday effect is based on the principle that a material can be magnetized by a circularly polarized electromagnetic wave. a) When the wave is left circularly polarized, the magnetization is oriented in a direction opposite to the wave propagation. b) Conversely, when the wave is right circularly polarized, the magnetization is oriented in the same direction as the wave propagation.

In this paper, we demonstrate, by manipulating light at the nanoscale, that we can generate a reversed IFE, i.e., an IFE whose magnetization is oriented in the direction of light propagation for a left circular polarization and oriented in the opposite direction to the propagation of light for a right circular polarization, as opposed to a so-called "classical" IFE. To do so, we have designed, thanks to an evolutionary algorithm, a plasmonic nanostructure allowing us to manipulate the polarization of the light around this structure locally. In particular, the created nano-object allows us to generate a left ellipticity of the light in the near-field for an excitation by a right circularly polarized light and, conversely, a right local ellipticity for an excitation by a left circularly polarized light. Also, due to the optical field enhancement capability of the plasmonic nanostructure, we show that this reversed IFE is very efficient, allowing the generation of a 6 mT magnetic field for the light power considered in this study. Finally, thanks to the possibility of inverse design algorithms to optimize several parameters simultaneously, the plasmonic objects developed here possess chirality properties allowing this reversed IFE to generate a single helicity of light, the mirror structure creating a non-zero magnetic field for the reverse helicity.

The results presented here are important for several reasons. First, we demonstrate for the first time that we can reverse the symmetry of an inverse Faraday effect thanks to the perfect control of the light in the near-field of a plasmonic structure. Then, the plasmonic approach being the only current technique allowing to create of confined, intense, and ultra-high magnetic fields, the possibility to generate a reversed IFE would help to answer the endless question on the nature of the all-optical-switching of magnetic layers in rare earth-transition metal alloy materials, in particular, if it is due to an inverse Faraday effect or to a magnetic dichroism effect in the considered layers.[37] Finally, the chiral character of this new physical phenomenon is particularly significant because it implies that this approach allows generating optically, even with incoherent unpolarized light, intense, ultrafast magnetic fields at the nanoscale, and always oriented in the same direction.

## 2. Results and discussions

Here, using a genetic algorithm (GA),[38-40] we optimized a plasmonic nanostructure made in a thin gold layer of 30 nm thickness deposited on a glass substrate (Figure 2). The nanostructure consists of a 2D matrix of 10x10 elements, each made up of metal or air and measuring 28 nm, for a total size of 280x280 nm$^2$. These dimensions were chosen for ease of fabrication using lithography techniques, and the corners of the nanostructure were smoothed to avoid non-physical effects generated by numerical approaches (see Figure S1). We used the finite difference time domain (FDTD) method in Lumerical software to simulate the structure, solving Maxwell's equations in space and time. The simulation window had dimensions of 750x750x900 nm$^3$ and was surrounded by a perfectly matched layer (PML) boundary to avoid parasitic reflection. The computational space was discretized using several meshes, with a coarser non-uniform mesh of 4 to 16 nm for the external parts of the simulation, a finer mesh of 4 nm for the central part of 288x288x36 nm$^3$ containing the nanostructures, and an even finer mesh of 1 nm for the part where the magnetic field was calculated of 140x140x32 nm$^3$. The nanostructures were excited by a pulsed plane wave of duration 5.3 fs with a peak power of 10$^{10}$ W/cm$^2$, corresponding to an energy density of 53 µJ/cm$^2$, which was chosen to be well below the material's threshold.[41,42] The convergence of the simulation was obtained when the energy inside the calculation window was lower than $10^{-5}$ of the initial injected energy. The gold properties in these simulations were taken from the textbook values of Johnson and Christy. To create the desired reversed IFE, each generation of the GA optimization was composed of 200 elements, and for each element, two simulations were performed using right and left circular polarizations. The associated drift currents were calculated (Equation (1)), and Biot and Savart's law (Equation (2)) estimated the Z-components of the fields $B_{RCP}$ and $B_{LCP}$, generated under these two excitation conditions at the center of the nanostructure in X, Y, and Z (symbolized by the blue star in Figure 2), where $B_{RCP}$ corresponds to the magnetic field under a right circular polarization of excitation, and $B_{LCP}$ corresponds to the left circular polarization. We then chose, as a GA optimization function, to maximize the difference abs($B_{RCP}$)-abs($B_{LCP}$) with $B_{RCP}$ being negative. This optimization function's purpose was to generate a chiral response of the plasmonic structure and create a reversed IFE. The evolution from a generation N to N+1, N being the number of the generation, was achieved by keeping the 200 best structures from generations 1 to N and breeding them to produce half of the elements of the next generation, with the other half being constituted of mutated elements with a mutation rate of 10%.

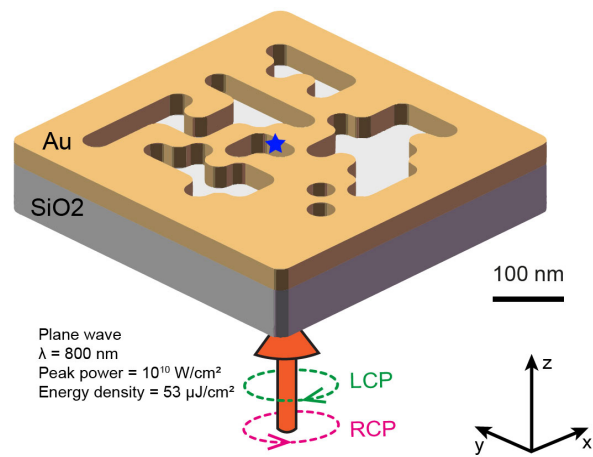

**Figure 2:** Optimized structure and excitation conditions. An example of a GA-optimized structure, consisting of a 30 nm thick gold layer and designed for excitation by a plane wave with right- or left-circular polarization at a wavelength of 800 nm and an excitation energy density of 53 µJ/cm$^2$, is shown. The resulting magnetic field **B** generated under these two excitation conditions is assessed at the center of the gold nanostructure in the X, Y, and Z directions, as represented by the blue star in the structure.

After 73 generations (Figure S2), an optimized structure is found, represented in Figure 3a. The optical response of this nano-object is shown in Figures 3b and c. These figures show the distribution of the electric field enhancement in an XY plane at the Z surface of the gold layer for excitation by a left and right circularly polarized light, respectively. As one can see, only the right circular polarization strongly increases the electric field locally, in good agreement with the optimization function given to the GA (the chirality objective). From these electric field distributions and via Equation (1), the drift currents present in the metal are calculated and plotted in Figures 3d and e in the same plane as the electric field distribution. These currents exist only for one polarization of the light, the right circular polarization, in good agreement with the chirality objective as well. Finally, these currents have a clockwise symmetry while the right circular polarization used to excite the nano-structure has an anti-clockwise symmetry, also in good agreement with the GA selection and opposite to a "classical" IFE.

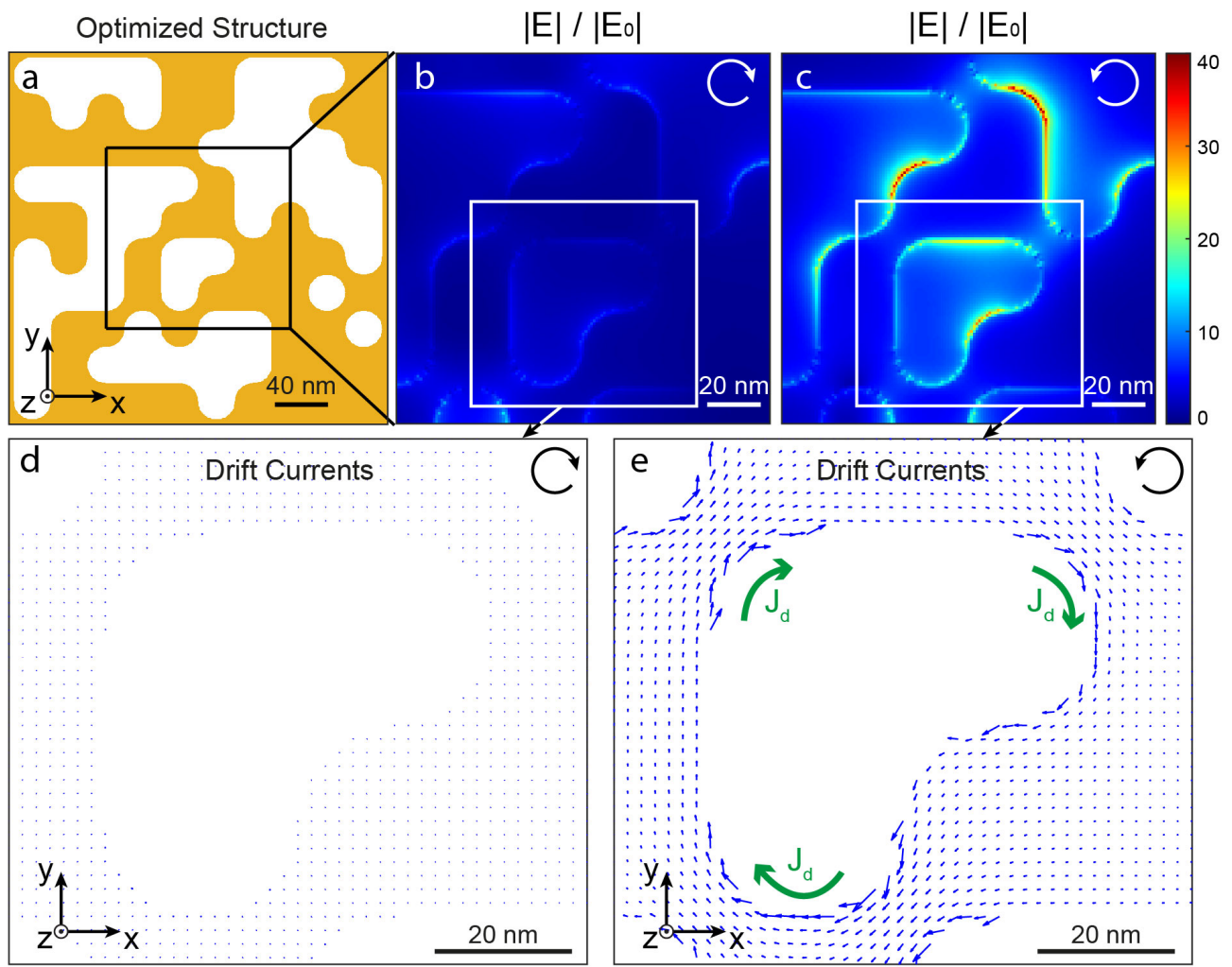

**Figure 3:** a) Schematic of the GA-optimized nanostructure in an XY plane. Spatial distribution of the electric field normalized by the incident wave at the Z-surface of the optimized nanostructure and in an XY plane symbolized by the black square in a) for excitation by a b) left and c) right circularly polarized plane wave. Spatial distribution of drift currents at the Z-surface of the optimized nanostructure in an XY plane symbolized by the white squares in b) and c), for excitation by a d) left and e) right circularly polarized light.

From these currents and via Equation (2), it is then possible to calculate the stationary magnetic field generated by the optimized nanostructure. Figure 4 represents these distributions of the magnetic field oriented along Z at the center of the optimized plasmonic nanostructure. In particular, Figures 4b and c represent the magnetic field in an XY plane at the center Z of the optimized structure described in Figure 4a, derived from the drift currents described in Figures 3d and e, respectively, for left and right circularly polarized excitation. Once again, we can see that a magnetic field is generated for only one excitation polarization, the right circular one, demonstrating the chiral effect expected by the optimization function. Also, and as expected in this study, the magnetic field generated by the structure of Figure 4a is negative and thus oriented towards negative Z, in contrast to a "classical" IFE, which magnetizes the matter in the direction of light propagation for this type of polarization of excitation. This result shows that manipulating light at the nanoscale allows the generation of a reversed IFE, which was not believed possible until today. Also, and as requested in the optimization code, when the mirror structure is used (Figure 4d), a magnetic field is generated only for the opposite polarization, i.e., left circular (Figure 4e, f), in this case, positive and thus always opposite to the magnetization by a "classical" IFE.

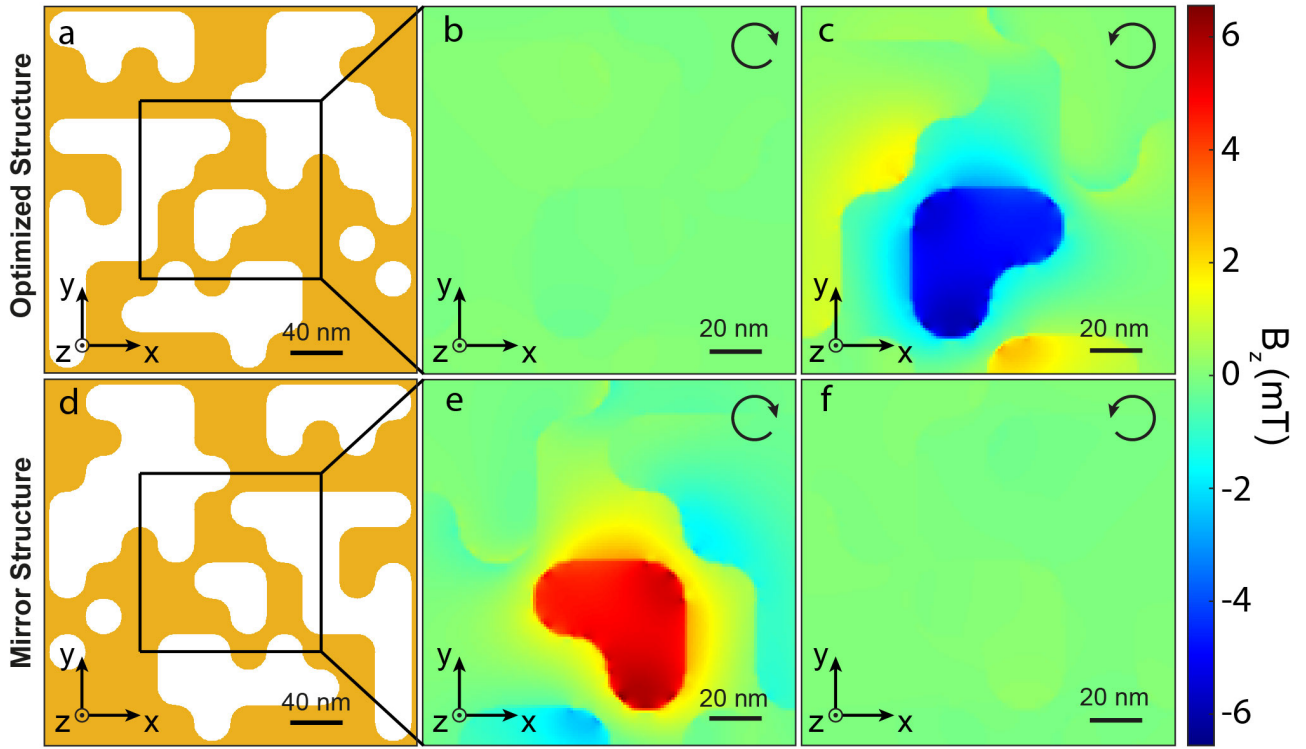

**Figure 4:** Magnetic response of the optimized plasmonic nanostructure. a) This figure shows again a schematic of the GA-optimized structure in an XY plane. b) and c) The spatial distribution of the magnetic field oriented along Z generated in the Z-center of the structure shown in a) is displayed for left and right circular polarizations of excitation, respectively. d) The figure also includes a schematic of the mirror structure displayed in a) in an XY plane. e) and f) Spatial distributions of the **B**-field oriented along Z and generated in the Z-center of the mirror structure shown in d) for left and right circular polarizations of excitation, respectively. The black arrows indicate the incoming polarizations.

The ability of optical nanostructures to manipulate light and its characteristics in the near field is at the origin of this new physical effect. Indeed, it has been demonstrated that the manipulation of electromagnetic fields in the near field enables, for instance, to control local densities of states,[43-45] radiation patterns,[46] chirality densities,[47,48] or some nonlinear effects.[49] Moreover, from Equation (1), we see that to generate drift currents at the origin of the IFE, a necessary condition must be met. The light incident on the metal must carry a certain degree of ellipticity so that the product **E.E*** is not zero. In particular, in the case of a classical IFE and as mentioned previously, a right circular or elliptical polarization will generate a stationary magnetic field oriented in the same direction as the propagation of the light. In contrast, a left circular polarization will create it opposite to the propagation direction

(Figure 1),[12,14] while, a linear polarization will not generate a drift current and, therefore, no magnetic field. Here, we use the unique ability of plasmonic nanostructures to locally manipulate the spin densities of light, or in other words, its local helicity, to generate a reversed IFE. The equation describing the electric spin density of light is:

$$\boldsymbol{s} = \frac{1}{|\boldsymbol{E_0}|^2} Im(\boldsymbol{E}^* \times \boldsymbol{E}) \qquad (3)$$

With $E_0$ the electric field of the incoming light.

The spin density is a vectorial physical quantity that describes the polarization state of light in a given plane. This density can take positive or negative values corresponding to right or left elliptical polarizations. In particular, a positive spin density in our reference system corresponds to a right helicity, a negative spin density corresponds to a left helicity, and a zero density corresponds to a linear polarization. In the far field, the spin density can only take values between -1 and 1, -1 corresponding to a left circular polarization and 1 to a right circular polarization (Figure 5a and b). However, in the near field, once normalized by the incident intensity |$E_0$|², the spin density can take much larger values due to the increase of the fields, leading to the concept of super-circular light[18] by analogy with super-chiral light.[50] Therefore, since an elliptical or circular polarization is required to generate drift currents via Equation (1), creating locally non-zero spin densities allows the generation of an IFE in a plasmonic nanostructure. In the present case, the capability of our optimized nanostructure to generate a reversed IFE comes from the fact that when the latter is excited by a right circularly polarized plane wave, it generates locally a hot spot of left elliptically polarized spin density (Figure 5b, d), thus creating a magnetization opposite to a "classical" IFE (Figure 4c). Moreover, the chirality properties of our nanostructure stem from the fact that for the two excitation polarizations, right and left circular, only one generates a non-zero spin density, the right circular polarization (Figure 5c, d). These two behaviors, of chirality and manipulation of the local spin density, arise from the constructive and destructive interference of light existing in the optimized nanostructure. Likewise, the mirror structure to the optimized nano-object behaves similarly for, this time, a left circular excitation polarization. In particular, this mirror structure creates a local hot spot of positive spin density (right elliptical, Figure 5e) for excitation by a left circularly polarized wave (negative spin density, Figure 5a). Therefore, as the optimized structure, the mirror object creates a non-zero spin density only for one excitation polarization, the left circular polarization (Figure 5e, f). From this spin density study, the chiral behavior of these nanostructures, together with the reversed IFE, are thus perfectly explained. Finally, the strong magnetic field generated by these plasmonic structures, considering the low fluence applied (5.3 µJ/cm²), can be understood by the strong increase of the spin density with a super-circular light up to -70 and 70 depending on the ellipticity (Figure 5d, e), which generates strong drift currents.

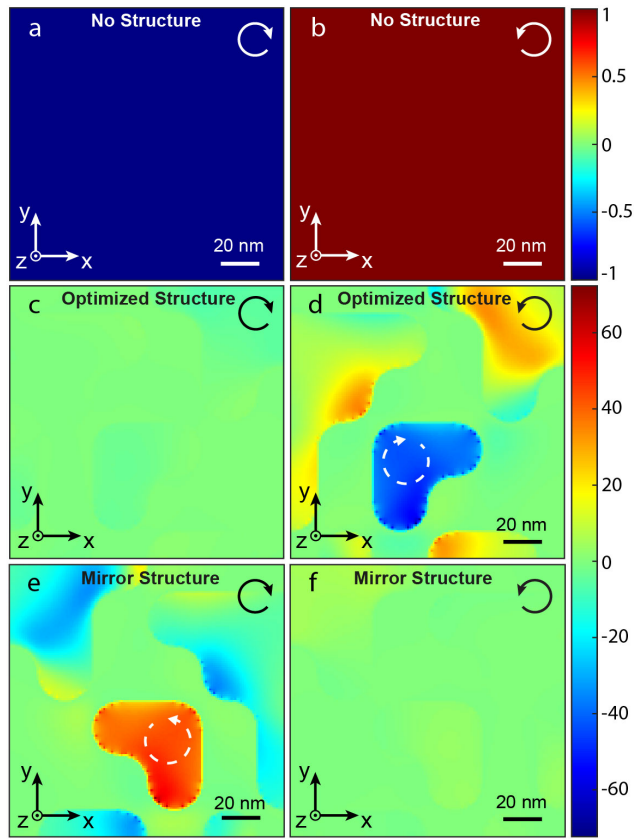

**Figure 5:** Distribution of spin densities. a) And b) show the spin density for left and right circularly polarized plane waves, respectively. Panels c) and d) show the local spin density in the Z-center of the optimized structure for left and right circular polarization of excitation. e) And f) show the local spin density in the Z-center of the mirror structure for excitation by a left and right circularly polarized wave, respectively. The full line white and black arrows indicate the incoming polarizations, the white dashed arrows show the direction of local elliptical polarizations.

## 3. Conclusion

In conclusion, we have demonstrated for the first time that the magneto-optical process of the inverse Faraday effect can be a reversed mechanism compared to the so-called "classical" IFE. This new physical behavior results from manipulating the polarization of light at the nanoscale. Using an inverse design algorithm based on natural selection, we demonstrated the generation of a local elliptical polarization whose helicity was opposite to the exciting circular polarization incident on the nanostructure, generating, in turn, this reversed IFE. Also, we demonstrated that this optimized structure exhibited chirality properties, generating a magnetic field only for one helicity of the light, thanks again to manipulating its polarization at the nanoscale. Indeed, the genetic optimization of this structure selected it to generate a non-zero spin density for one helicity of the light only, the mirror structure generating a magnetization for the opposite handedness. Finally, thanks to the super-circular light created by these structures, a concept similar to super-chiral light, which describes an elliptical polarization of light where the optical fields are strongly enhanced, the generated **B**-field has an amplitude of 6 mT. This magnetic field is one of the most intense created at these scales and by IFE, given the low fluence of 5.3 µJ/cm$^2$ exciting the optimized nanostructure. The results presented here are significant for several reasons. First, we have demonstrated for the first time that the magneto-optical process of IFE can be reversed. Also, these results pave the way for a reversed magnetization of matter by IFE for unpolarized light excitation. Finally, IFE by plasmonic nanostructures is the only approach to generating ultrafast magnetic field pulses at the nanoscale.[14,16] The ability to induce a magnetic field that is defined by the structure for unpolarized excitation would have applications in the manipulation of magnetic processes such as skyrmion manipulation, ultrafast magnetic modulation, magnetic trapping, spin currents, or spin precession, with, for example, ultrafast data writing as a direct application.

**Acknowledgements**

The authors declare no conflict of interest.
We acknowledge the financial support from the Agence national de la Recherche (ANR-20-CE09-0031-01), from the Institut de Physique du CNRS (Tremplin@INP 2020) and the China Scholarship Council.

# A Reversed Inverse Faraday Effect

Ye Mou, Xingyu Yang, Bruno Gallas, and Mathieu Mivelle*

Sorbonne Université, CNRS, Institut des NanoSciences de Paris, INSP, F-75005 Paris, France

*Corresponding author: mathieu.mivelle@sorbonne-universite.fr

A list of the main content:
Simulation parameters
Supporting figures S1 to S3

**Simulation parameters:**
The simulations carried out in this study were done by the finite difference time domain (FDTD) method performed on the commercial software Lumerical from Ansys. This method solves Maxwell's equations in space and time using a finite difference technique. Indeed, the FDTD method solves these equations on a discrete spatial and temporal grid. The dimensions of the 3D computational window for the simulations were 750x750x900 nm$^3$. The boundary conditions of this window are made of a perfectly matched layer (PML), avoiding any parasitic reflection inside the calculation window. Several meshes are used for the discretization of the computational space, a coarser non-uniform mesh of 4 to 16 nm for the external unstructured parts, of the simulation, a finer mesh of 4 nm for a central nanostructured part of 288x288x36 nm$^3$ in X, Y, and Z, respectively containing the nanostructures, and an even finer mesh of 1 nm for the part where the drift currents and magnetic field are calculated of 140x140x32 nm$^3$ in X, Y, and Z (Figure S1c). The choice of this mesh size for the central part is chosen because convergence in the amplitude of the magnetic field is observed starting from this mesh size. The excitation of the nanostructures is performed by a pulsed plane wave of duration 5.3 fs spectrally centered at a wavelength of 800 nm. The peak power of the pulse is 10$^{10}$ W/cm$^2$, which corresponds to an energy slightly lower than 0.048 pJ applied to the plasmonic nanostructures (energy density of 53 µJ/cm$^2$). The convergence of the simulation was obtained when the energy inside the calculation window was lower than 10$^{-5}$ of the initial injected energy. The textbook values of Johnson and Christy were used for the gold properties in these simulations.

**Supporting figures:**

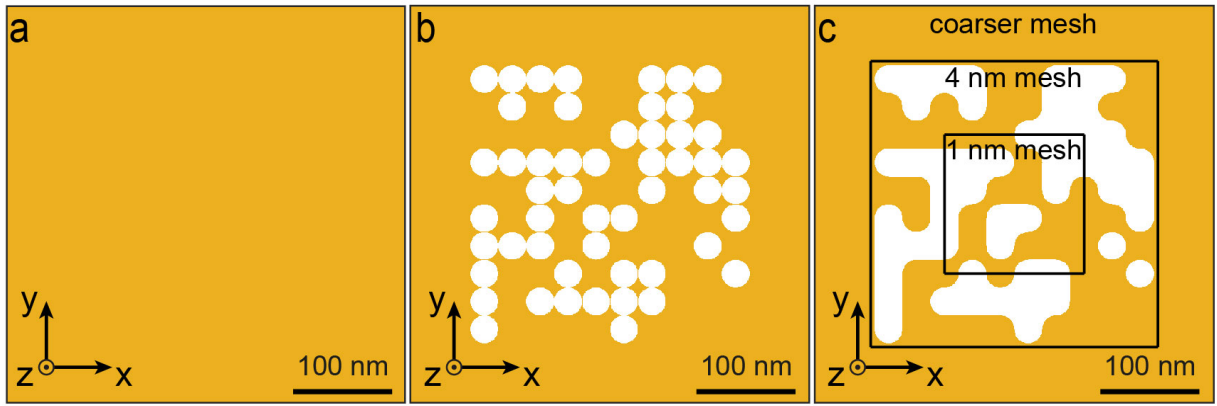

**Figure S1:** Construction of the elements constituting each generation of the genetic algorithm. Inside a) a uniform gold layer of 30 nm thickness, b) holes are made according to a binary matrix playing the role of the DNA in the evolutionary process. c) The obtained structure is then smoothed to avoid all the roughnesses not experimentally feasible and generating non-physical effects. The different mesh areas are shown in c.

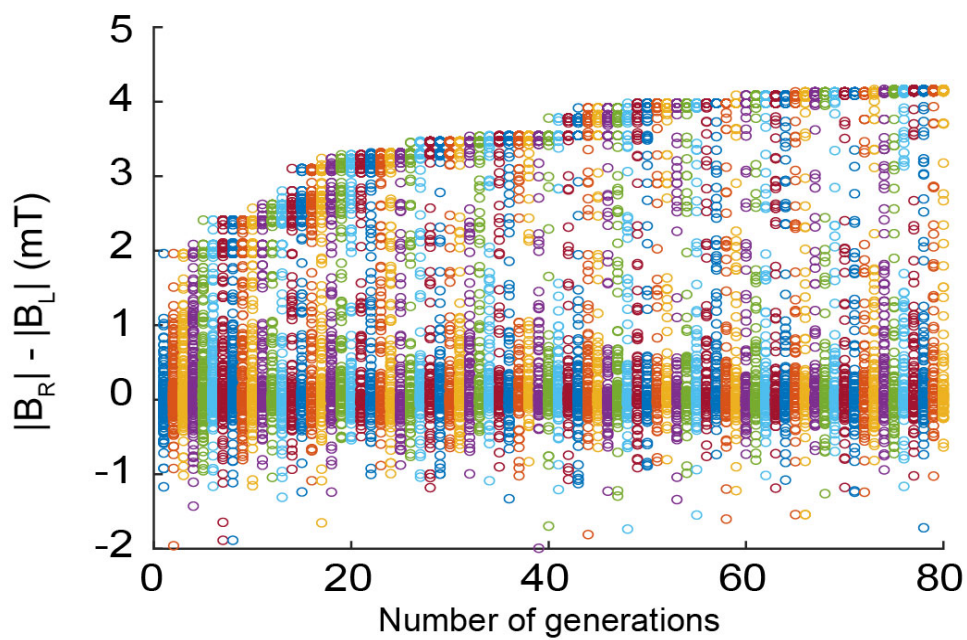

**Figure S2:** Evolutionary process. Evolution during the different generations of the optimization function consisting in maximizing the difference abs($B_{RCP}$)-abs($B_{LCP}$) with $B_{RCP}$ being negative, $B_{RCP}$ and $B_{LCP}$ are the **B**-fields created by a right or left circular polarization, respectively. Each generation consists of 200 structures. The optimized structure appears after 73 generations.

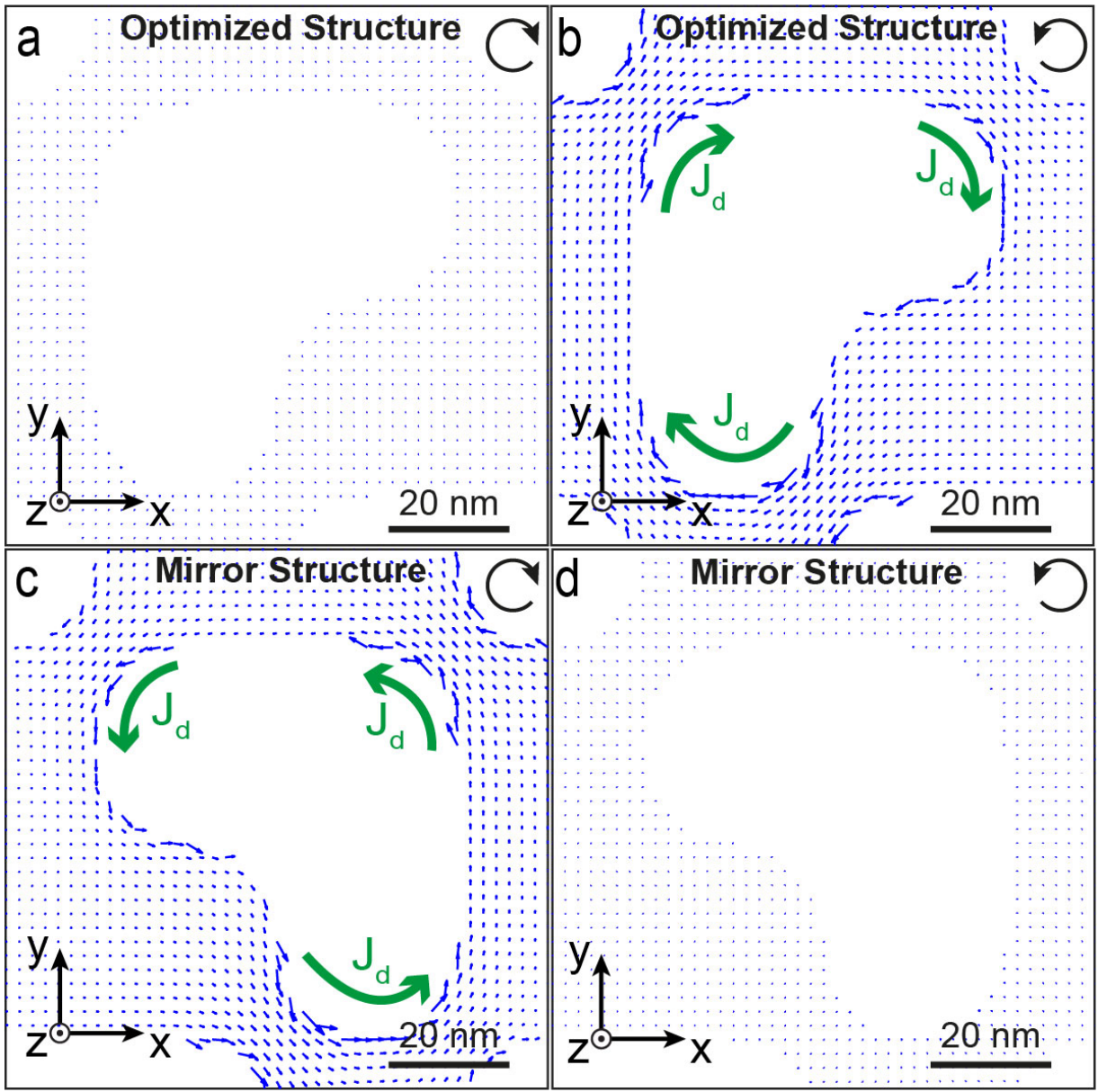

**Figure S3:** Distribution of drift currents in an XY plane. a) and b) Spatial distributions of drift currents at the surface (1 nm below the upper edge of the gold layer) of the optimized structure for the left and right circular polarizations of excitation, respectively. c) and d) Spatial distributions of drift currents at the surface of the mirror structure for the left and right circular polarizations of excitation, respectively. The black arrows indicate the incoming polarizations.